\documentclass[%
 reprint,
 amsmath,amssymb,
aps,
prb,
]{revtex4-1}

\usepackage{graphicx}
\usepackage{dcolumn}
\usepackage{bm}

\begin{document}
	

\preprint{APS/123-QED}

\title{Magnetic Correlations in the Quasi-2D Semiconducting 
Ferromagnet CrSiTe$_3$}

\author{T.J.~Williams}
 \email{williamstj@ornl.gov}
\author{A.A.~Aczel}
\author{M.D.~Lumsden}
\author{S.E.~Nagler}
\author{M.B.~Stone}
\affiliation{Quantum Condensed Matter Division,
    Neutron Sciences Directorate,
    Oak Ridge National Lab,
    Oak Ridge, TN, 37831, USA}

\author{J.-Q.~Yan}
\author{D.~Mandrus}
\affiliation{Materials Science \& Technology Division,
    Physical Sciences Directorate,
    Oak Ridge National Lab,
    Oak Ridge, TN, 37831, USA}
\affiliation{Department of Materials Science \& Engineering,
 	University of Tennessee,
 	Knoxville, TN, 37996, USA}

\date{\today}

\begin{abstract}
Intrinsic, two-dimensional ferromagnetic semiconductors are an important 
class of materials for overcoming the limitations of dilute magnetic 
semiconductors for spintronics applications.  CrSiTe$_3$ is a particularly 
interesting member of this class, since it can likely be exfoliated down to 
single layers, where T$_C$ is predicted to increase dramatically.  
Establishing the nature of the magnetism in the bulk is a necessary precursor 
to understanding the magnetic behavior in thin film samples and the possible 
applications of this material.  In this work, we use elastic and inelastic 
neutron scattering to measure the magnetic properties of single crystalline 
CrSiTe$_3$.  We find that there is a very small single ion anisotropy favoring 
magnetic ordering along the $c$-axis and that the measured spin waves fit well 
to a model where the moments are only weakly coupled along that direction.  
Finally, we find that both static and dynamic correlations persist within the 
$ab$-plane up to at least 300~K, strong evidence of this material's 
two-dimensional characteristics that are relevant for future studies on thin 
film and monolayer samples. 

\begin{description}
\item[PACS numbers]{75.30.Ds, 75.40.-s, 75.50.Pp, 85.75.-d}
\end{description}
\end{abstract}

\maketitle

\section{\label{sec:level1}Introduction}

The isolation of single layers of carbon atoms, in the form of graphene, 
resulted in the first physical realization of a true two-dimensional (2D) 
crystal leading to considerable study and revealing new 
physics~\cite{Geim_07}.  The interest in graphene was not only fundamental, 
but also applied; potential device applications, such as spintronic 
devices, of such two-dimensional materials were recognized soon after the 
initial discovery~\cite{Han_14}.  Device applications of graphene are limited 
by its intrinsic properties, for instance the lack of a band gap, and the 
search for 2D crystals soon extended to other materials.  Spintronics requires 
semiconductors that possess ferromagnetic properties, particularly those that 
can be manipulated into one- or two-dimensional forms, so much work has 
focused on dilute magnetic 
semiconductors~\cite{Zutic_04,MacDonald_05,Dietl_10,Deng_11}, which offer a 
large range of magnetic properties.  These compounds have been the subject of 
much debate since their properties are highly dependent on epitaxial growth 
conditions, dopant distributions and even how the dopants are incorporated 
into the overall band structure~\cite{Dietl_10}.  In an effort to overcome 
these limitations, there has been an emergence of research into the much less 
common spintronic candidate: intrinsically ferromagnetic 
semiconductors~\cite{Rogado_05,Li_14}.  The compound CrSiTe$_3$ belongs to 
this class, having an electronic band gap of 0.38~meV and a ferromagnetic 
transition at T$_C$~=~33~K. 

Much of the recent work on CrSiTe$_3$ has been in the form of theoretical 
calculations focused on the band structure and semiconducting properties, 
particularly calculations of the monolayer properties.  Generalized gradient 
approximation (GGA) calculations on bulk CrSiTe$_3$ have calculated a bulk 
band gap of 0.6~meV, close to the experimental value, with the gap formed by 
the splitting of the Cr and Te levels~\cite{Lebegue_13}.  These calculations 
correctly predict a ferromagnetic ground state that induces a slight spin 
polarization in the Si and Te atoms~\cite{Lebegue_13}.  When the spins were 
assumed to be Ising-like and aligned along the $c$-axis, local density 
approximation (LDA) calculations predicted a nearest-neighbor exchange 
$J$~=~-0.58~meV and transition temperature T$_C$~=~23~K~\cite{Siberchot_96}.  
Experimentally, the structure has been determined in earlier 
work~\cite{Ouvrard_88}. Hexagonal planes of Cr$^{3+}$ ($S=3/2$) atoms are 
stacked along the c-axis, with each atom octahedrally-coordinated by Te. The 
Te-Te bond lengths are $\approx$~3.15~\AA~in the $ab$-plane and 
$\approx$~3.48~\AA~out of the plane, suggesting a very small octahedral 
distortion~\cite{Ouvrard_88,Casto_15}.  Previous neutron measurements also 
found a magnetic transition at 32.1~K corresponding to the Cr$^{3+}$ moments 
ordering ferromagnetically along the $c$-axis~\cite{Carteaux_95}.  These 
measurements found a spin gap of nearly 6~meV, which was taken as evidence for 
Ising behavior.  However, the inelastic measurements suffered from low 
statistics, which casts doubt on the accuracy of the determined magnetic 
dynamics.  Additionally, they do not measure in multiple Brillouin Zones, 
making it possible that they have misidentified phonon modes as magnons.  This 
offers an explanation as to why their measurements find three spin wave 
branches, a conclusion that is inconsistent with linear spin wave theory based 
on the known magnetic structure.  Additionally, the claim that the spins are 
Ising-like cannot be conclusively derived from these measurements for the same 
reason.  Other experimental work cited in that work does not contain 
evidence of Ising spins, since the cited magnetostriction 
measurements~\cite{Herpin_68} are only dependent on the exchange anisotropy, 
while measurements of the single-ion anisotropy show it to be quite 
small~\cite{Abragam_69,deJongh_90}.  Finally, the behavior of the magnetic 
correlations above T$_C$ were only discussed qualitatively, leaving much room 
for a more accurate neutron scattering study to determine the magnetic 
correlations.

The resurgent interest in studying CrSiTe$_3$ has been driven by its 
applicability to spintronics.  This has been prompted by the speculation that 
it may be possible to exfoliate the material down to single 
layers~\cite{Li_14}, which has been predicted to have the desirable effect of 
increasing the band gap to $\approx$~0.59~eV~\cite{Li_14} and the Curie 
temperature to $\approx$~92~K~\cite{Li_14,Sivadas_15}.  These features have 
quickly created the need to understand the electronic and magnetic properties 
of CrSiTe$_3$.

\section{\label{sec:level2}Experimental Details}

\begin{figure}[tbh]
\begin{center}
\includegraphics[angle=0,width=\columnwidth]{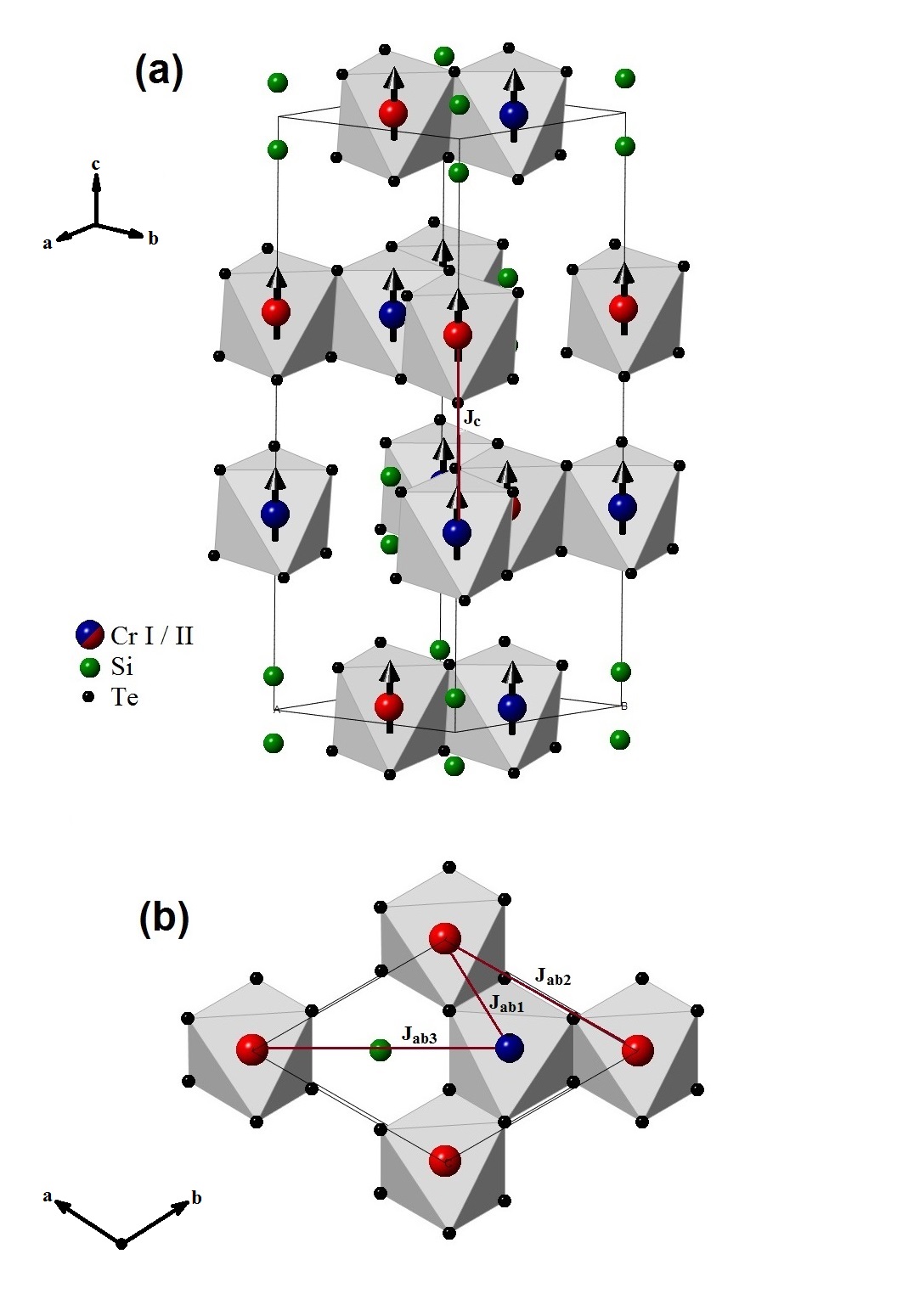}
\caption{\label{structure} (color online) (a) The crystal structure of 
CrSiTe$_3$.  The Cr$^{3+}$ ions form hexagonal arrangements in the 
$ab$-plane, which are then stacked along the $c$-axis in an $ABC$-type 
stacking.  This creates two magnetically-inequivalent Cr sites, shown in red 
and blue.  The Cr-I site (blue) has a Cr-II atom (red) above and a 	Si$_2$ 
pair (green) below, while it is reversed for the Cr-II site.  Below 	
T$_C$~=~33.2~K, the Cr $S$~=~3/2 spins align ferromagnetically along the 
$c$-axis, due to a single-ion anisotropy, $D_z$.  The nearest-neighbour 
exchanges out to 8~\AA~are shown in the figure, which includes 3 in-plane 
exchanges, $J_{ab1}$, $J_{ab2}$ and $J_{ab3}$, and one out-of-plane exchange, 
$J_{c}$.  (b) The structure of a single layer is shown, highlighting the 
different exchange interactions in this plane.}
\end{center}
\end{figure}

CrSiTe$_3$ single crystals were grown using a self-flux technique, as 
previously reported~\cite{Casto_15}.  CrSiTe$_3$ is rhombohedral, 
crystallizing in the space group $R$-3, which was confirmed with x-ray 
diffraction measurements on the samples used in this study.  The crystal 
structure is shown in Fig.~\ref{structure}.  Susceptibility measurements 
showed a ferromagnetic transition at T$_{C}$~=~33(1)~K~\cite{Casto_15}.  The 
large $c/a$ ratio and the reasonably accessible magnetic transition make this 
a good candidate to study for its application as a two-dimensional spintronic 
material.  In order to fully characterize the nature of the magnetic 
correlations, including their lower dimensional properties, neutron scattering 
measurements were performed at the HB-3 and CG-4C (CTAX) triple axis 
spectrometers of the High-Flux Isotope Reactor, as well as the SEQUOIA 
time-of-flight spectrometer at the Spallation Neutron Source of the Oak Ridge 
National Laboratory.

Five single crystals of total mass 4.1g and a mosaic of 2.25$^{\circ}$ were 
coaligned in the [$H$~0~$L$] scattering plane for use in the CTAX 
experiment. This array of single crystals was also used for the SEQUOIA 
experiment, while the largest single crystal (mass~$\approx$~1.2g) was used 
for the HB-3 experiment, also aligned in the [H~0~L] scattering plane.  The 
HB-3 measurements were performed in a closed-cycle refrigerator with a base 
temperature of 4.0~K using a fixed final energy of 14.7~meV.  PG (002) 
monochromator and analyzer crystals were used with PG filters, and the 
collimation was 48'-40'-40'-120'.  The SEQUOIA measurements were also 
performed in a closed-cycle refrigerator using fixed incident energies of 30 
and 65~meV.  The crystals were rotated in the [H~0~L] plane in 1$^{\circ}$ 
steps over an 83$^{\circ}$ range.  The CTAX measurements were 
performed in a He-4 cryostat with a base temperature of 1.5~K using fixed 
final energies of 3 and 5~meV.  This experiment used a PG002 monochromator, a 
Be filter and collimation settings of guide-open-80'-open.

\section{\label{sec:level3}Elastic Neutron Scattering}

\begin{figure}[tbh]
\begin{center}
\includegraphics[angle=0,width=\columnwidth]{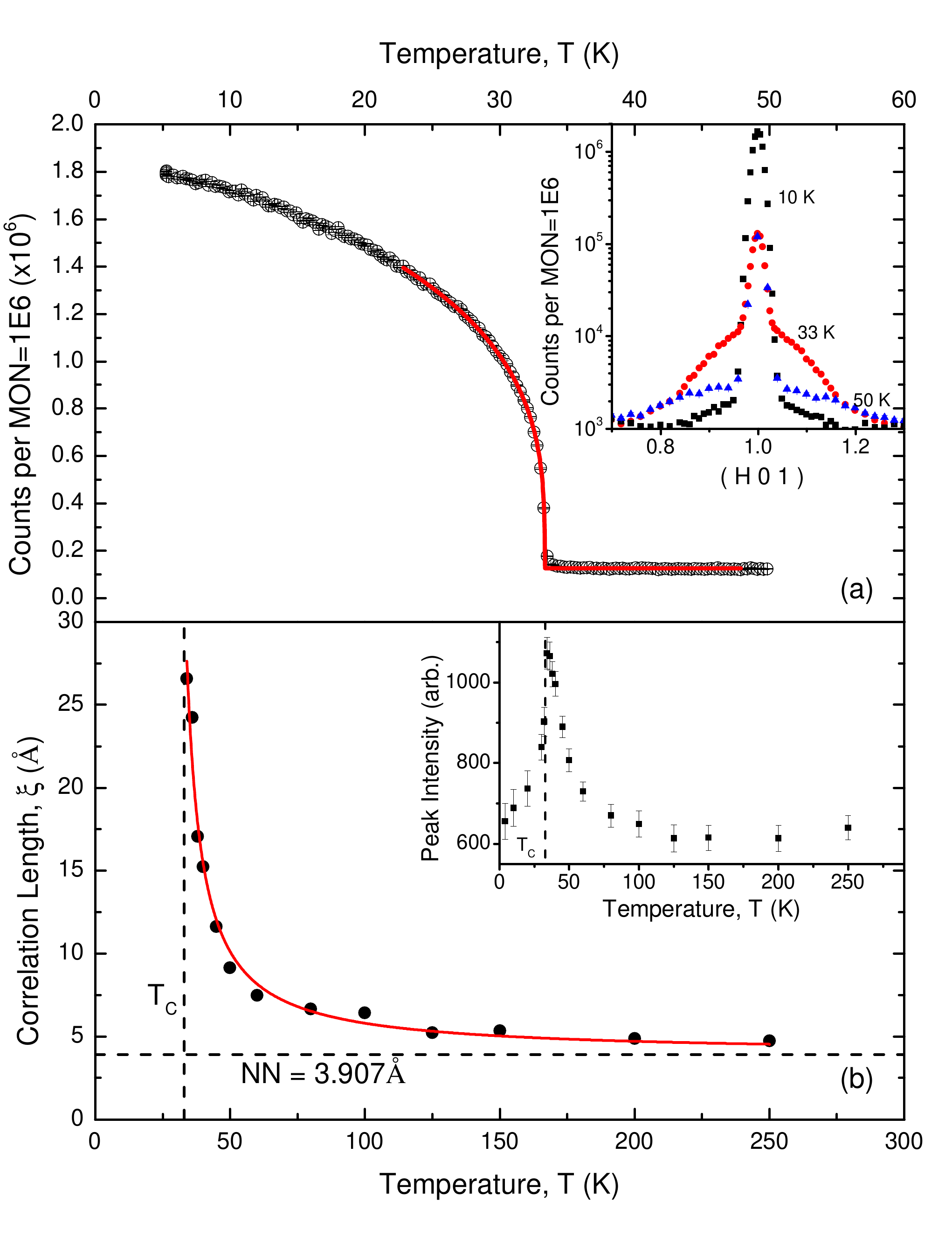}
\caption{\label{hb3} (color online) (a) The temperature dependence of the 
measured intensity at the (1~0~1) Bragg peak.  This peak is both nuclear and 
magnetic, but the magnetic intensity is larger by approximately a factor of 
10.  A fit to this curve gives a transition temperature of T$_C$~=~33.2(1)~K 
and a critical exponent $\beta$~=~0.151(12).  (inset) Scans through the 
(1~0~1) Bragg peak along $H$ at T~=~10~K (black squares), 33~K (red circles) 
and 50~K (blue triangles), plotted on a logarithmic scale.  This shows the 
sharp increase in the intensity at (1~0~1), corresponding to the 3D order, 
while the diffuse scattering, corresponding to the 2D correlations, is most 
intense at the transition.  (b) The temperature dependence of the in-plane 
correlation length, determined from the 2-axis scans described in the text.  
The line is a guide to the eye.  We  note that the in-plane correlation length 
is greater than the nearest-neighbor Cr-Cr distance at all temperatures 
measured, suggesting that the in-plane magnetic correlations are important 
well above the bulk ordering temperature.}
\end{center}
\end{figure}

\begin{figure*}[tbh]
\begin{center}
\includegraphics[angle=0,width=7in]{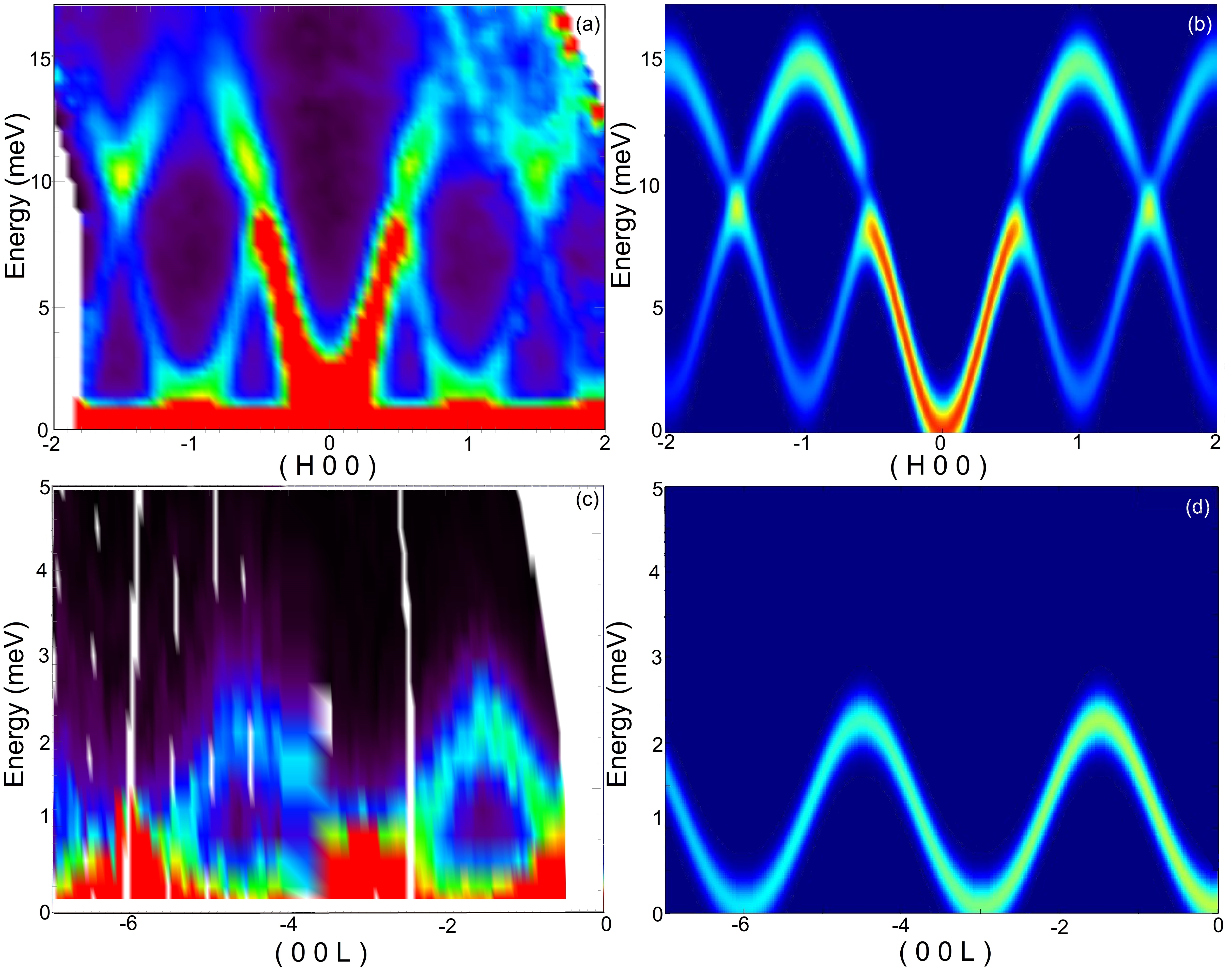}
\caption{\label{combined} (color online) The panels on the left show the spin 
waves of CrSiTe$_3$ measured on the SEQUOIA spectrometer at T~=~10~K, while 
the panels on the right show the calculated spin wave dispersions, along the 
($H$~0~0) direction (panels (a) and (b)) and the (0~0~$L$) direction (panels 
(c) and (d)).  The calculated patterns use the model and exchange parameters 
described in the text, producing very good agreement with the experimental 
data.}
\end{center}
\end{figure*}

Fig.~\ref{hb3}(a) shows the temperature dependence of the 
(1~0~1) Bragg peak, which is both a nuclear and a magnetic peak, but the 
magnetic contribution is approximately 10 times larger at 4~K.  The red line 
is a fit to a critical exponent, which yields a transition temperature of 
33.2(1)~K and a critical exponent, $\beta$~=~0.151(12).  This is close to the 
value expected for a two dimensional transition ($\beta_{2D,Ising}$~=~0.125), 
and well below the values expected for a three dimensional transition 
($\beta_{3D,Ising}$~=~0.326 and $\beta_{3D,Heisenberg}$~=~0.367).  The low 
value of critical exponent is likely a consequence of strong two-dimensional 
correlations in this material.  The long-range order below 33.2~K is, however, 
three-dimensional in nature, with the spins aligned ferromagnetically along 
the $c$-axis direction.  This was confirmed by measuring 39 magnetic peaks in 
the [$H$~0~$L$] scattering plane, whose intensities below T$_C$ were 
consistent with $c$-axis ordering.  This is in agreement with previous 
work~\cite{Carteaux_95}.

In addition to the magnetic diffraction peaks that emerge below T$_C$, diffuse 
scattering develops around the Bragg peaks, with an intensity that has a 
maximum at T$_C$, shown in the inset to Fig.~\ref{hb3}(a).  This diffuse 
scattering arises due to the in-plane correlations that are present, 
particularly above the ordering temperature.  In order to study the 
magnetic correlations within the $ab$-plane as a function of temperature, a 
series of two-axis measurements were performed on the HB-3 instrument.  This 
was done by measuring along $H$ about the wavevector (1~0~0.545), chosen such 
that the $c$-axis was parallel to the final neutron wavevector, 
$k_f$~\cite{Birgeneau_70,Als-Nielsen_75}.  The analyzer was then removed to 
integrate over various momentum transfers, but the $\hat{c} \parallel k_f$ 
orientation means that the component of the momentum transfer that varies with 
final energy is constrained to be along the $L$-direction.  This provides an 
accurate measure of the energy-integrated spin response for two-dimensional 
correlations within the $ab$-plane. 

The two-axis measurements are peaked at $H$~=~1 for all temperatures 
measured and we find that the peak intensity is largest at T$_C$, which is 
shown in the inset to Fig.~\ref{hb3}(b).  The width of the peak can be used to 
calculate the in-plane correlation length, which is plotted in 
Fig.~\ref{hb3}(b).  As expected, the correlation length peaks at T$_C$, but 
decays as the temperature increases.  This experiment did not have the 
required temperature stability or density of data points near T$_C$ to perform 
an accurate measurement of the critical exponent, $\nu$, or to observe the 
expected 2D to 3D crossover, which typically occurs within 0.01\% to 0.1\% of 
T$_C$~\cite{Taroni_08}.  However, we do observe that the correlation length 
remains larger than the nearest-neighbor distance at all temperatures 
measured, up to 250~K.  This suggests that short-range correlations between 
the moments in the $ab$-plane exist well above the ordering transition.  

\section{\label{sec:level4}Spin Wave Measurements}

\begin{figure*}[tb]
\begin{center}
\includegraphics[angle=0,width=7in]{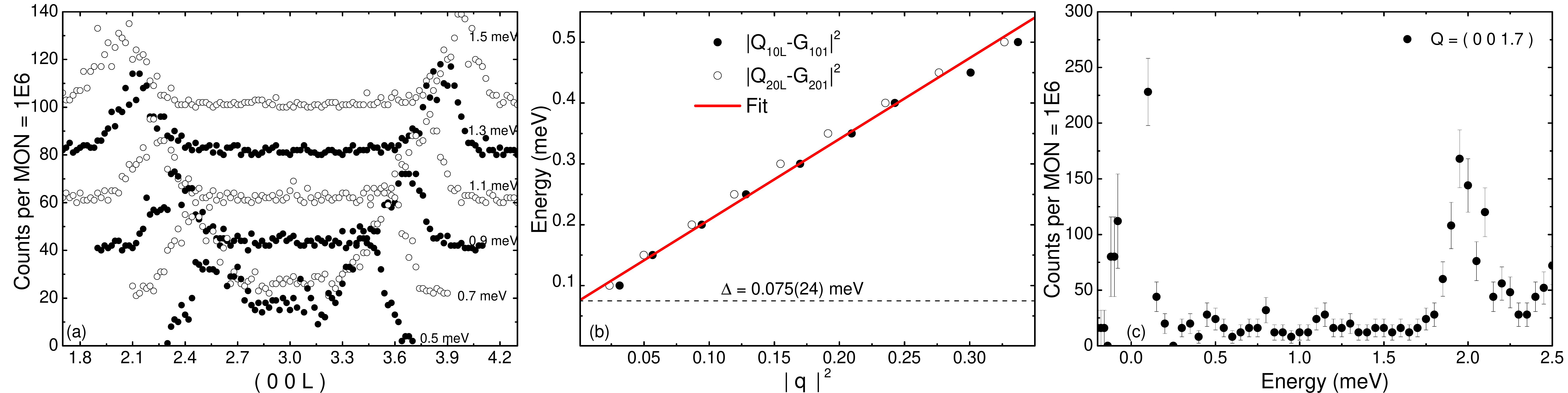}
\caption{\label{ctax_fits} (color online) (a) The constant-$E$ scans measured 
on the Cold Triple-Axis (CTAX) Spectrometer at 1.5~K along (0~0~L), which are 
offset along the y-axis for clarity.  These scans focus on the region near the 
zone center (0~0~3), where a spin gap is expected in the spectrum.  By 
assuming that the energy of the spin waves, $E \propto \left|q\right|^2$ in 
the long wavelength limit, we can extract an approximate measure of the gap, 
despite it being within the resolution of the instrument. (b) The 
$q$-dependence of the spin waves near the zone center is plotted, showing 
$q^2$-dependence.  A fit to these values gives a value for the gap 
$\Delta = 0.075(24)$~meV.  (c) A constant-$\vec{Q}$ measurement at 
$\vec{Q}$~=~(~1.7~0~0) near the zone boundary, (0~0~1.5), allows for a 
determination of the value of the out-of-plane coupling, 
$J_c$~=~-0.730(96)~meV.}
\end{center}
\end{figure*}

Fig.~\ref{combined}(a) and (c) show the spin wave dispersions measured on the 
SEQUOIA spectrometer, with an incident energy E$_i$~=~30~meV.  Panel (c) 
shows the dispersion along (0~0~$L$) with integration ranges of 0.4 reciprocal 
lattice units (r.l.u) along $H$ and $K$.  This shows that the dispersion is 
very weak along the $c$-axis, and that the minimum in the spin wave dispersion 
is below 1~meV at the zone center, (0~0~-3).  Since the dispersion along $L$ 
is relatively weak, it was observed that varying the integration range along 
$L$ had little effect on the plot of the spin wave spectrum.  Therefore, the 
spectrum shown in Fig.~\ref{combined}(a) was constructed by integrating over 
the full $L$-range, $-12 \le L \le 12$, while using an integration range of 
0.2~r.l.u. along $K$.  This shows two spin wave bands, one peaking at 
$\approx$~8~meV and the higher band peaking at $\approx$~15~meV, which meet at 
the zone boundaries.  The data along ($H$~0~0) using a smaller integration 
range for $L$ showed identical dispersions, supporting the conclusion that the 
spin waves are close to two-dimensional.  The spin waves were modeled with a 
Hamiltonian given by:

\begin{equation}
\begin{split}
H&= J_{ab1} \sum_{i}{\vec{S}_i \cdot \vec{S}_{i+1}} + 
J_{ab2} \sum_{i}{\vec{S}_i \cdot \vec{S}_{i+2}} \\
&+ J_{c} \sum_{i}{\vec{S}_i \cdot \vec{S}_{i+3}} +
J_{ab3} \sum_{i}{\vec{S}_i \cdot \vec{S}_{i+4}} - 
D_z \sum_{i}{(\vec{S}_i^z)^2} 
\label{eq1}
\end{split}
\end{equation}

\noindent
where the $J$'s are exchange constants between neighboring spins, $\vec{S}_i$, 
$D_z$ represents a single-ion anisotropy and the spins are assumed to be 
localized Cr$^{3+}$, $S=3/2$ moments, as shown in Fig.~\ref{structure}.  The 
very small value of the spin gap and the weak $c$-axis dispersion are both 
able to be accurately determined from the SEQUOIA data.  The magnetic 
anisotropy necessitates a non-zero spin gap, but the SEQUOIA measurements have 
shown that it is less than 1~meV.  This is consistent with magnetization data 
taken to characterize the sample, which observes very little magnetic 
hysteresis and a field of 1.5~T applied in the $ab$-plane (the hard 
magnetization direction) is required to saturate the magnetic moment.  
Additionally, DFT calculations based on purely van~der~Waals interactions have 
found that there is an entropy difference of $\approx$~20~$\mu$eV, favoring 
$c$-axis ordering, as compared to ordering within the 
$ab$-plane~\cite{Li_14,Zhuang_14}.  This is very close to the value obtained 
for the single-ion anisotropy, as discussed below.  The fits described below 
produce the simulations shown in Fig.~\ref{combined}(b) and (d).  These show 
very good agreement with the experimental data, but we do observe a 
discrepancy at the non-zero integer values of $H$.  At these points in the 
experimental data, we observe broadening of the optical mode and intensity 
shifts to the acoustic mode.  These features are not seen in the simulation, 
though the origin of this difference is unclear.  

In order to measure the magnitude of the spin gap and the dispersion out of 
the plane, measurements were performed on the cold neutron triple axis (CTAX) 
instrument at Oak Ridge National Laboratory.  The energy resolution of a 
constant-$Q$ scan was insufficient to separate the gapped mode from the 
elastic scattering.  To extract the spin gap value, a series of constant-$E$ 
scans was performed, shown in Fig.~\ref{ctax_fits}(a).  These scans focus on 
the region near the zone center (0~0~3), where a spin gap is expected in the 
spectrum.  By assuming that the energy of the spin waves, $E \propto 
\left|q\right|^2$ in the long wavelength limit, we can extract an approximate 
measure of the gap, despite it being within the resolution of the instrument.  
The data was fit in multiple zones, as shown in Fig~\ref{ctax_fits}(b), 
showing $q^2$-dependence.  This gave a value of the spin gap, 
$\Delta$~=~0.075(24)~meV, allowing a measurement of the single-ion anisotropy, 
$D_z = \Delta / 2 S$~=~0.0252(80)~meV.  The single-ion anisotropy arises from 
the small $c$-axis distortion in the Te octahedra surrounding the Cr$^{3+}$ 
ions.  This nearly perfect octahedral environment makes spin-orbit effects 
very weak and the Cr-Cr bond is a point of local centrosymmetry, ruling out 
Dyaloshinskii-Moriya interactions as an origin for the energy gap.  This small 
gap is consistent with theoretical predictions that the energy difference 
between $c$-axis and $ab$-plane ordering is less than 0.1~meV.  
Constant-$\vec{Q}$ measurements near the zone boundary, (0~0~1.5), shown in 
Fig.~\ref{ctax_fits}(c), allow us to determine the value of the out-of-plane 
coupling, $J_c$, since the energy of the lower spin wave branch along $L$ 
depends only on $D_z$ and $J_c$.  Using the value of $D_z$ determined above, 
we obtain $J_c$~=~-0.730(96)~meV.

These values were then fixed for the purposes of fitting the SEQUOIA data to 
obtain the in-plane exchange constants.  The fitting was performed using the 
fitting routines built into the Horace software package~\cite{Horace_15} which 
used the dispersions calculated by SpinW~\cite{Toth_14} as its model, 
comparing the intensities of the data and calculation over the entire range in the $\vec{Q}$-$E$ slice.  This allowed the in-plane exchange constants to be fit simultaneously, utilizing the two-dimensional dataset shown in Fig.~\ref{combined}(a).  The values of these three exchange constants obtained from fitting the data shown in Fig.~\ref{combined}(a) to Eq.~\ref{eq1}, as well as those determined from the CTAX measurements, are given in Table~\ref{exch_tbl}.

\begin{table}[htb]
\begin{center}
\begin{tabular}{|l|l|l|l|}
\hline
 Exchange & ~~~~~~Description & ~Distance~ & Value (meV)
  \\
\hline
~~~$J_{ab1}$ & ~~~1$^{st}$~NN in-plane & ~~3.907 \AA & ~~~-1.27(23) \\
~~~$J_{ab2}$ & ~~~2$^{nd}$~NN in-plane & ~~6.768 \AA & ~~~-0.10(50) \\
~~~$J_{ab3}$ & ~~~3$^{rd}$~NN in-plane & ~~7.814 \AA & ~~-0.285(73) \\
~~~~$J_{c}$ & ~1$^{st}$~NN out-of-plane & ~~6.852 \AA & ~~-0.730(96) \\
~~~~$D_z$ & Single Ion Anisotropy &  & ~~0.0252(80) \\
\hline
\end{tabular}
\end{center}
\caption[]{The exchange constants obtained from fitting the inelastic 
measurements using Eq.~\ref{eq1}.  To properly describe the spin waves, it 
was necessary to use couplings up to 8~\AA, which requires 3 in-plane 
interactions and one out-of-plane.  Consistent with the quasi-2D nature of 
the material, the value of $J_c$ gives a very small dispersion along the 
$c$-axis and, as expected from the very small octahedral distortion, the value 
of the single-ion anisotropy, $D_z$, is very close to zero.}
\label{exch_tbl}
\end{table}

\begin{figure*}[tb]
\begin{center}
\includegraphics[angle=0,width=7in]{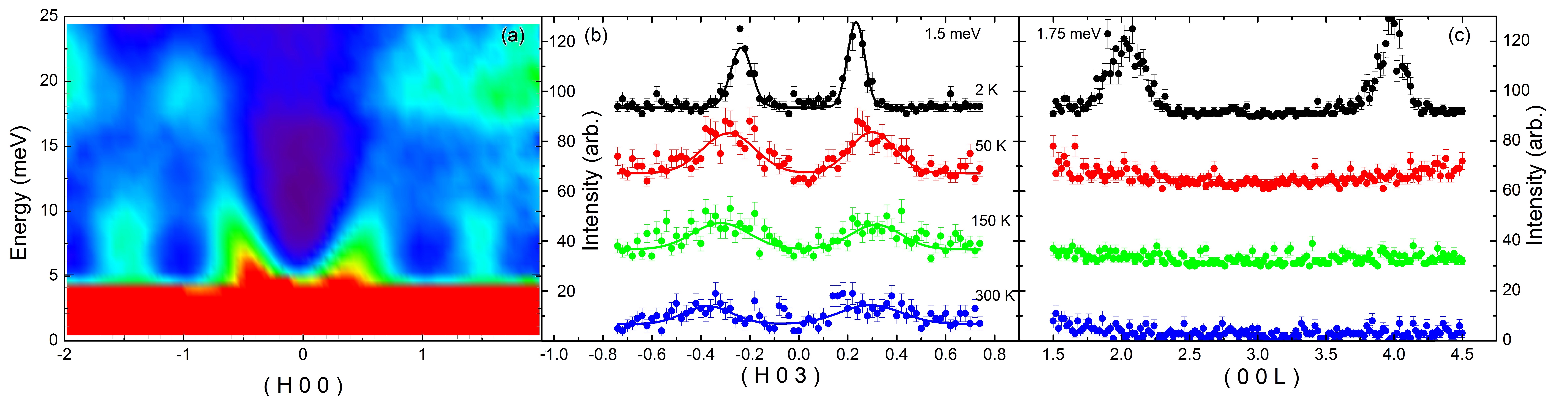}
\caption{\label{ctax} (color online) The temperature dependence of spin 
dynamics.  (a) The inelastic neutron spectrum measured at T~=~40~K on the 
SEQUOIA spectrometer.  We see that broadened spin wave features remain, even 
above the transition, while there are no such excitations along the $L$ 
direction.  This is a clear signature of the two-dimensionality of this 
material.  The feature at 20~meV is extrinsic to the sample, verified by 
performing an empty can measurement.  (b) The in-plane spin dynamics along 
(H~0~3) at various temperatures, measured on CTAX and offset for clarity.  
Sharp spin waves are observed below the transition, but dynamic correlations 
exist at all temperatures measured.  The high temperature features are 
reminiscent of the spin waves seen below T$_C$, but significantly broadened.  
The lines are guides to the eye. (c) In contrast, there are no dynamic 
correlations along the $L$ direction above T$_C$ (also shown offset for 
clarity).  This suggests that three-dimensional correlations only exist below 
the Curie temperature, while the in-plane dynamic correlations persist up to 
at least 300~K.}
\end{center}
\end{figure*}

As expected for a quasi-two-dimensional system, the nearest neighbor 
in-plane coupling, $J_{ab1}$, is the dominant interaction and is 
ferromagnetic, as are the other in-plane exchange constants.  While $J_{ab1}$ 
is only slightly larger than $J_{c}$, there are 3 nearest neighbors in the 
$ab$-plane compared to only one along the $c$-axis.  This makes the in-plane 
exchange more than 5 times larger than the out-of-plane coupling.  The 
ferromagnetic origin of the in-plane terms is likely due to superexchange 
mediated by the Te ions, shown in Fig.~\ref{structure}(b).  The Cr$^{3+}$ ion 
direct exchange is antiferromagnetic, though the large nearest-neighbor Cr-Cr 
distance of 3.987~$\textrm{\AA}$ should make this a weak 
effect~\cite{Colombet_83,Motida_70}.  Finally, the Cr octahedra are 
edge-sharing and the Cr-Te-Cr angle is 88.8$^{\circ}$, very close to 
perfectly orthogonal.  This arrangement of the ions would suggest that 
superexchange of the type described by the Goodenough-Kanamori 
rules~\cite{Goodenough_55,Goodenough_58,Kanamori_59} would be ferromagnetic, 
consistent with the observations of ferromagnetism in other layered 
Cr$^{3+}$ compounds with similar superexchange pathways~\cite{Motida_70}.  
First principles DFT calculations on monolayer CrSiTe$_3$ suggest that the 
2$^{nd}$ and 3$^{rd}$ nearest-neighbor in-plane interactions are Cr-Te-Te-Cr 
double-superexchange interactions~\cite{Sivadas_15}, leading to the small 
values we observe for $J_{ab2}$ and $J_{ab3}$.  Furthermore, the observation 
that $\left| J_{ab2} \right| < \left| J_{ab3} \right|$ was also predicted by 
the DFT calculations. There are two double-superexchange pathways that 
contribute to $J_{ab2}$, one of which is ferromagnetic and the other 
antiferromagnetic, making it a very weak exchange~\cite{Sivadas_15}.

The neutron scattering measurements are also consistent with the theoretical 
predictions of an enhanced transition temperature in CrGeTe$_3$ and 
CrSnTe$_3$, as well as monolayer 
CrSiTe$_3$~\cite{Li_14,Sivadas_15,Zhuang_14}.  Despite the spins being only 
weakly Ising-like, the easy-axis spin anisotropy created by the imperfect Te 
octahedra still allows for ordering in two-dimensions, such as when the 
materials are reduced to monolayers.  The loss of the out-of plane coupling is 
compensated by an increased in-plane ferromagnetic exchange, as 
predicted by several theoretical studies that compare bulk CrSiTe$_3$ to its 
monolayer version and the chemically-substituted CrGeTe$_3$ and 
CrSnTe$_3$~\cite{Li_14,Sivadas_15,Zhuang_14}.  The increased ferromagnetic 
exchange within the $ab$-plane arises due to the increase of the 
nearest-neighbor Cr-Cr distance in all of these cases.  This serves to reduce 
the Cr-Cr direct exchange, which is antiferromagnetic, while pushing the 
Cr-Te-Cr bond angle closer to 90$^{\circ}$.  This increases the effect of the 
ferromagnetic superexchange, resulting in the increased values of 
T$_C$~\cite{Motida_70}.

Inelastic neutron scattering measurements performed above the ferromagnetic 
transition are shown in Fig.~\ref{ctax}.  Measurements at 40~K along the 
($H$~0~0) direction are presented in Fig.~\ref{ctax}(a).  This data indicates 
that the spin waves are still present within the planes, but have been 
significantly broadened.  The minima in the spin waves seen at (1~0~0) and 
(2~0~0) in Fig.~\ref{combined}(a) have disappeared, as they arose from the 
weak dispersion along the $L$-direction.  This indicates that there is no 
dispersion along $L$ above T$_C$, and these broadened spin waves are only 
present within the $ab$-planes at these temperatures.  To track the 
temperature evolution of the spin waves, constant-$E$ measurements were 
performed along ($H$~0~0) at $E$~=~1.75~meV and (0~0~$L$) at $E$~=~1.5~meV as 
a function of temperature.  These are shown in Fig.~\ref{ctax}(b) and (c), 
respectively.  We see that the spin waves along $L$ disappear abruptly above 
$T_C$, but that the broadened spin waves exist at the same $\vec{Q}$ along 
($H$~0~0) well above the ordering transition, up to at least 300~K.

\section{\label{sec:level5}Conclusions}

We have measured CrSiTe$_3$ using elastic and inelastic neutron scattering. 
These measurements observe bulk ferromagnetism below T$_C$~=~33.2(1)~K and the 
critical exponent of the neutron scattering intensity was found to be 
$\beta$~=~0.151(2), close to the value expected for a two-dimensional 
system.  The magnetic Bragg peaks also exhibited a diffuse component, the 
intensity of which peaks at T$_C$, indicating two-dimensional ferromagnetic 
correlations are present in the $ab$-plane above the ferromagnetic 
transition.  To characterize these correlations, two-axis measurements were 
performed, which provided an effective quantitative measure of the in-plane 
correlation length.  As expected, the correlation length diverged at T$_C$ but 
was still larger than the nearest-neighbor Cr-Cr distance at all temperatures 
measured, up to 250~K.  These measurements suggest that while CrSiTe$_3$ only 
orders in three dimensions below 33~K, there are strong two-dimensional static 
correlations that persist up to at least room temperature.

In contrast to the previous assumption of Ising spins, the spin wave 
measurements suggest that the spins are very Heisenberg-like, with a very 
small spin gap of 0.075(24)~meV due to a small single-ion anisotropy.  X-ray 
diffraction measurements~\cite{Casto_15} and DFT calculations~\cite{Zhuang_14} 
find evidence of van~der~Waals interactions creating a very small octahedral 
distortion along the $c$-axis, creating the anisotropy.  We find that the 
dominant interaction, $J_{ab1}$~=~-1.27(23)~meV, is likely due to the 
superexchange mediated by the Te ions.  The 2$^{nd}$ and 3$^{rd}$ 
nearest-neighbor in-plane interactions are much weaker, likely a result of 
double-superexchange interactions.  

Above T$_C$, no spin waves are present along the $L$ direction, but they 
persist in a broadened form along the $H$ direction.  Measurements at 40~K 
along ($H$~0~0) showed broadened spin waves consistent with dynamics confined 
to the $ab$-planes, a clear indication of two-dimensional behavior, while 
temperature-dependent measurements show an inelastic signal that is visible up 
to 300~K.  The latter finding suggests that there are dynamic in-plane 
correlations that persist up to at least room temperature as well.  
Considering these results in conjunction with the neutron diffraction 
measurements, we find that both static and dynamic in-plane magnetic 
correlations exist up to at least 300~K, ten times T$_C$.  This illustrates 
the importance of two-dimensional correlations in these materials and may 
suggest that similar physics is relevant in the isostructural compounds 
CrGeTe$_3$ and CrSnTe$_3$, as well as for monolayer CrSiTe$_3$, where it has 
been predicted that increased separation or complete decoupling of the 
hexagonal Cr layers puts an increased emphasis on the two-dimensional 
correlations.  Taken together with the dramatic increase in the magnetic 
transition temperature driven by a separation of the intralayer Cr atoms in 
these materials, this suggests that spintronic devices~\cite{Han_14} that make 
use of these semiconductors on a substrate that enhances the magnetic 
character through strain or other effects may be capable of supporting 
intrinsic ferromagnetism at temperatures approaching, or even surpassing, room 
temperature.

\section{\label{sec:level6}Acknowledgments}

We acknowledge instrument support from S.~Chi, T.~Hong and J.~Niedziela.  This 
research at ORNL's High Flux Isotope Reactor and Spallation Neutron Source was 
sponsored by the Scientific User Facilities Division, Office of Basic Energy 
Sciences, US Department of Energy.  T.J.W. acknowledges support from the 
Wigner Fellowship program at Oak Ridge National Laboratory.  D.G.M and J.-Q.Y. 
acknowledge support from NSF DMR 1410428.

\end{document}